\documentclass{nature}
\usepackage{color}
\usepackage{amssymb}
\usepackage{acronym}
\usepackage{amsmath}
\usepackage{hyperref}
\usepackage{graphicx}
\usepackage{lineno}

\hypersetup{
    pdfnewwindow=true,      
    colorlinks=true,       
    linkcolor=blue,          
    citecolor=blue,        
    filecolor=blue,      
    urlcolor=blue           
}

\newcommand*{\addd}[1]{\textcolor{black}{#1}}
\newcommand*{\adddd}[1]{\textcolor{black}{#1}}

\makeatletter
\let\saved@includegraphics\includegraphics
\AtBeginDocument{\let\includegraphics\saved@includegraphics}
\renewenvironment*{figure}{\@float{figure}}{\end@float}
\makeatother

\title{Eccentricity Estimate for Black Hole Mergers with Numerical Relativity Simulations}
\author{V. Gayathri$^{1}$, J. Healy$^{2}$, J. Lange$^{3,4,2}$, B. O'Brien$^{1}$, M. Szczepa\'nczyk$^{1}$, Imre Bartos$^{1}$, M. Campanelli$^2$, S. Klimenko$^{1}$, C.O. Lousto$^{2}$, R. O'Shaughnessy$^{2}$}

\begin{document}

\maketitle

\begin{affiliations}
\item Department of Physics, University of Florida, PO Box 118440, Gainesville, FL 32611-8440, USA
\item Center for Computational Relativity and Gravitation, Rochester Institute of Technology, Rochester, NY 14623, USA
\item Department  of  Physics,  University  of  Texas,  Austin,  TX  78712,  USA
\item Institute of Computational and Experimental Research in Mathematics, Brown University, Rhode Island 02903, USA
\end{affiliations}

\begin{abstract}
\addd{The origin of black hole mergers discovered by the LIGO\cite{2015CQGra..32g4001L} and  Virgo\cite{2015CQGra..32b4001A} gravitational-wave  observatories is currently unknown. 
GW190521\cite{GW190521discovery,GW190521_properties} is the heaviest black hole merger detected so far. Its observed high mass and possible spin-induced orbital precession could arise from the binary having formed following a close encounter. 
An observational signature of close encounters is} eccentric binary orbit\cite{2018ApJ...860....5G,2018PhRvD..97j3014S,2018PhRvD..97j3014S,samsing2020active};  
however, this feature is currently difficult to identify due to the lack of suitable gravitational waveforms. No eccentric merger has been previously found\cite{Abbott_eBBH}.  
Here we report 611 numerical relativity simulations covering the full eccentricity range \addd{and an estimation approach} to probe the eccentricity of \addd{mergers}. Our set of simulations corresponds to $\sim 10^5$ waveforms, comparable to the number used in gravitational wave searches, albeit with coarser mass-ratio and spin resolution. \addd{We applied our approach to GW190521 and found that} it is the most consistent with a highly eccentric ($e=0.69^{+0.17}_{-0.22}$; 90\% credible level) 
merger within our set of waveforms. \adddd{This interpretation is supported over a non-eccentric merger with $>10$ Odds ratio if $\gtrsim10\%$ of GW190521-like mergers are highly eccentric.} Detectable orbital eccentricity would be evidence against an isolated binary origin, \addd{which is otherwise difficult to rule} out based on observed mass and spin.\cite{2018PhRvD..98h4036G,2020A&A...635A..97B} 
%
\end{abstract}
Most black holes that are born in a binary system will orbit their companion for up to billions of years. Their initial orbital eccentricity will diminish over this time through the emission of gravitational waves. Black holes that are born with large initial natal kicks\cite{2005ApJ...618..845G} can substantially reduce the inspiral time, but even then only negligible eccentricity is expected at gravitational-wave frequencies LIGO-Virgo are sensitive to. On the other hand, binaries formed through gravitational capture in chance encounters can form with small orbital radii and high initial eccentricity, leaving insufficient time for the binary to lose its eccentricity before reaching the short orbital periods detectable by LIGO-Virgo. Alternatively, the interaction of a binary with a nearby black hole can also increase its eccentricity\cite{2017ApJ...836...39S,2019ApJ...881...41L,2018ApJ...856..140H}. 

Despite its importance in probing the origin of black hole mergers, it is difficult to identify orbital eccentricity through observations. Eccentricity expands the degrees of freedom of gravitational waveforms, making standard template-based searches problematic to carry out in practice. Eccentric orbits have a large dynamic range, with fast relativistic evolution near the pericenter and slower evolution closer to the apocenter, which presents complications both for waveform computations utilizing numerical relativity, and for those using post-Newtonian approximations\cite{Ireland:2019tao}. 

We carried out $611$ eccentric and $920$ non-eccentric numerical relativity simulations to produce a suite of binary merger gravitational waveforms that cover the full $e\in[0,1]$ eccentricity range, including non-spinning, aligned-/anti-aligned spin and spin-induced precessing waveforms (see Methods). These gravitational waveforms were then scaled to correspond to different black hole masses, providing about 100 scaled mass values for each simulation, corresponding to a total of approximately $10^5$ waveforms. Numerical simulations that include both eccentricity and precession were not previously available on this scale. We used these waveforms to self-consistently analyze the possibility and implications of GW190521 being an eccentric binary source. We estimated the binary properties of GW190521 using Bayesian parameter estimation through directly comparing our numerical relativity simulations with gravitational-wave strain data using the RIFT package\cite{gwastro-PENR-RIFT,gwastro-PENR-Methods-Lange}. To determine whether the best-matching gravitational waveform found by RIFT is a good match to the observed data in an absolute sense, we quantitatively tested this consistency using the model-independent waveform reconstruction algorithm {\it coherent WaveBurst}\cite{Salemi:2019uea} (cWB).  
Such a dual analysis has not been previously carried out, and is important in probing a so-far unexplored parameter space with a limited number of templates. 

\section{Waveform likelihood} 


Our numerical-relativity simulations were directly compared with gravitational-wave strain data using the RIFT package\cite{gwastro-PENR-RIFT,gwastro-PENR-Methods-Lange}. We assumed a Gaussian likelihood ${\cal L}(d|h)$, where $d$ is the recorded data and $h$ is the time-dependent strain of the incoming gravitational wave signal. Each numerical relativity simulation $\boldsymbol{\lambda}$ was described by the following intrinsic parameters: mass ratio $q$, black hole spins $\vec{S}_1$ and $\vec{S}_2$, eccentricity $e$ and mean anomaly (see Methods). The redshifted total mass $M_{\rm z}$ of the system was set by scaling the simulated waveform. For each $\boldsymbol{\lambda}$ we computed the marginal likelihood
\begin{equation}
{\cal L}_{\rm marg} = \int d\theta dM_{\rm z} p(\theta){\cal L}(d|h(\theta,\boldsymbol{\lambda},M_{\rm z}))\,, 
\label{eq:lmarg}
\end{equation}
where we marginalized via direct Monte Carlo integration\cite{gwastro-PENR-RIFT} over a set of seven extrinsic parameters (distance, time, two sky position angles, and three Euler angles describing the source orientation [(inclination, orbital phase, and polarization]) denoted with $\theta$ and the redshifted total mass $M_{\rm z}$. The probability $p(\theta,M_{\rm z})$ is the prior probability. We adopted a prior that is uniform in orientation, luminosity distance cubed and redshifted mass. 

\begin{figure}
\centering
  \includegraphics[angle=0,scale=0.58]{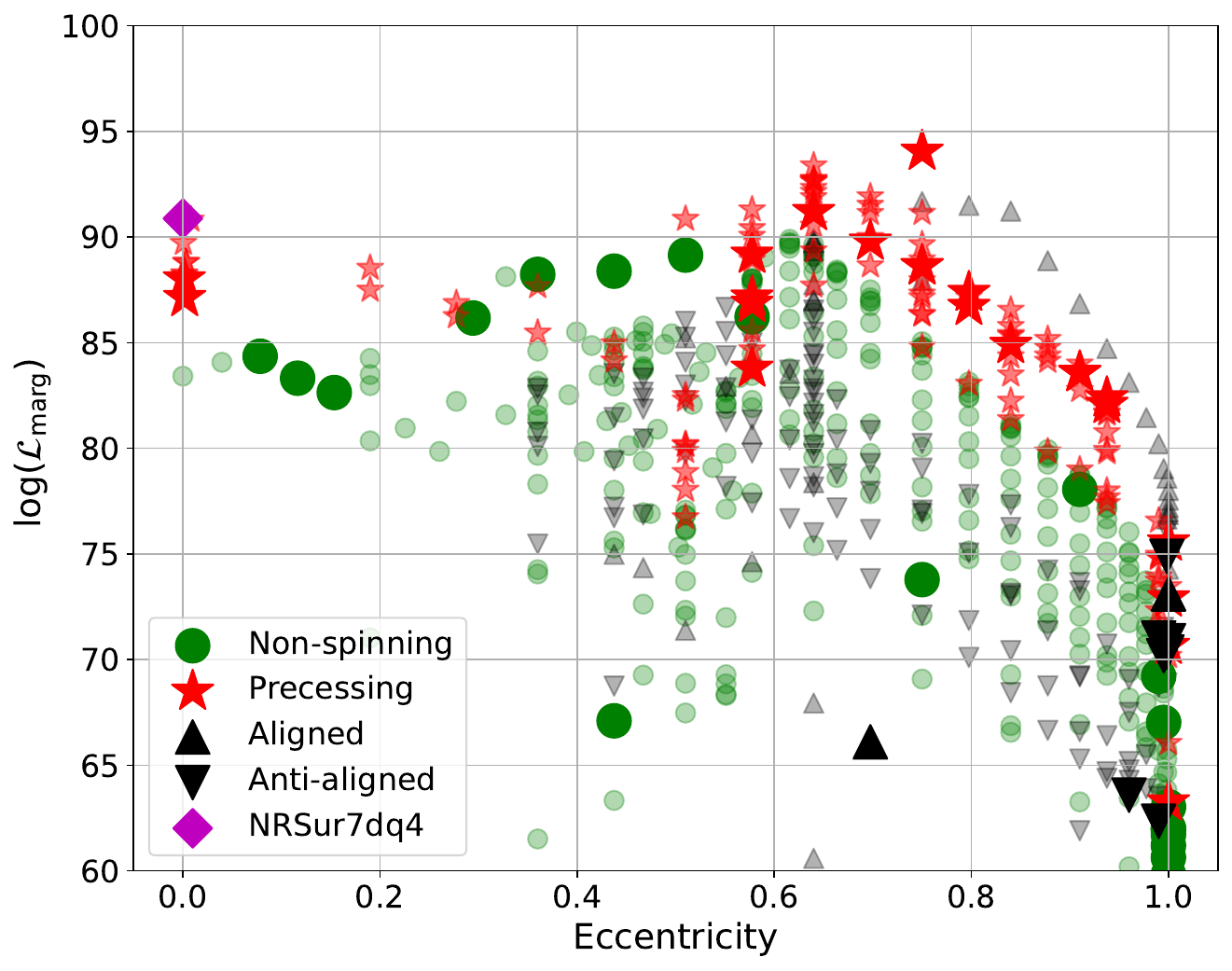}
      \caption{{\bf Marginalized likelihood as a function of eccentricity for our numerical relativity simulations explaining GW190521.} Each point corresponds to a separate numerical relativity simulation, and is categorized based on the black holes' spin (see legend). Categories include 'non-spinning' ($\chi_{\rm eff} = \chi_{\rm p}=0$), 'precessing' ($\chi_{\rm p}>0$), 'aligned' ($\chi_{\rm eff}>0$ \& $\chi_{\rm p}=0$) and 'anti-aligned' ($\chi_{\rm eff}<0$  \& $\chi_{\rm p}=0$). The simulations are further distinguished using a model-agnostic waveform consistency test. Large (small) marks correspond to simulations in which the reconstructed gravitational waveform is consistent (inconsistent) with the highest-likelihood waveform. We also include for comparison our likelihood estimate for the much more detailed set NRSur7dq4 of circular waveforms\cite{PhysRevResearch.1.033015} used by LIGO-Virgo\cite{GW190521discovery,GW190521_properties} (purple). Eccentricity for each simulation is defined at 10\,Hz assuming a binary mass of $50\,$M$_\odot$.
    \label{fig:LnL}}
\end{figure}

We computed the marginal likelihood value corresponding to each of our numerical relativity simulations using Eq. \ref{eq:lmarg}. The obtained values are shown in Fig. \ref{fig:LnL}. We grouped waveforms with respect to the black holes' spin and spin orientation. 

\section{Waveform consistency test} 


Our RIFT results select the best matching waveform to the data for each numerical relativity simulation through measuring the Gaussian likelihood of the waveform compared to the observed data. The accuracy of this likelihood test is, however, limited by the non-Gaussian detector noise. \addd{For instance, there could be non-Gaussian noise artifacts that mimic an eccentric waveform.}

\begin{figure}
    \includegraphics[scale=0.32]{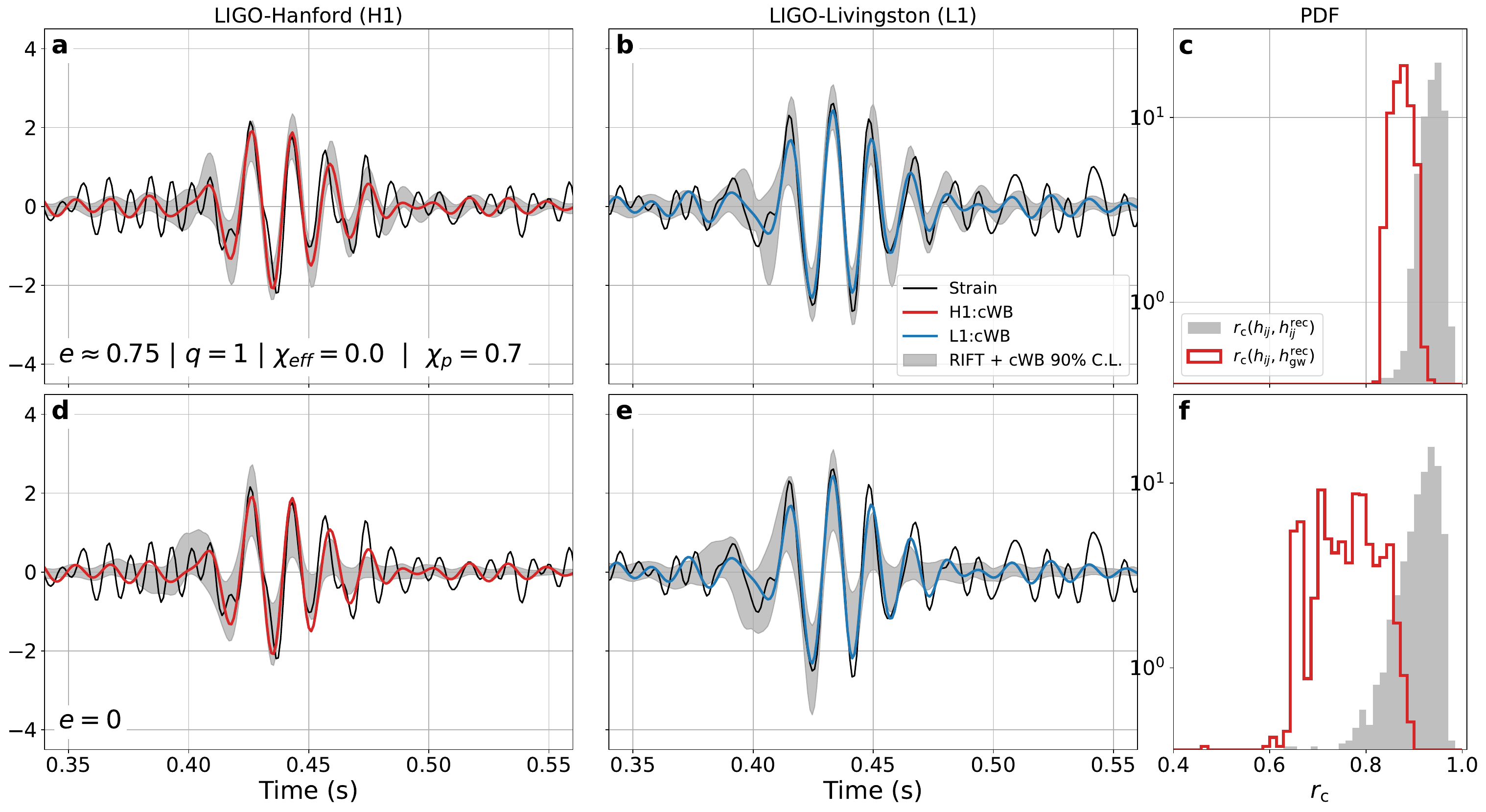}
    \caption{{\bf Reconstructed waveform of GW190521 and consistency test.} Top panels: high-eccentricity simulation of the present work; bottom panels:  circular NRSur7dq4 model\cite{PhysRevResearch.1.033015}; for the LIGO-Hanford (a,d) and LIGO-Livingston (b,e) detectors. The colored lines in (a,b,d,e) show the reconstructed waveform of GW190521 using cWB\cite{GW190521discovery} together with their 90\% confidence intervals (shaded regions; calculated over the weighted RIFT samples injected into the off-source data and reconstructed with cWB). Black lines show the recorded detector data after whitening with a band-pass filter ($28-128$\,Hz). Right column: distributions of $r_{\rm c}(h_{\it ij},h^{\rm rec}_{\it ij})$ and $r_{\rm c}(h_{\it ij},h^{\rm rec}_{\rm gw})$ (see legend). }
    \label{fig:time-domain}
\end{figure}

\addd{To account for such non-Gaussian effects,} we extended our likelihood test with a second step, called waveform consistency test. This test \addd{ measures whether the similarity of a simulated waveform $h_i$ to the measured gravitational wave signal could be due to detector noise}. 
For a numerical relativity waveform $i$, we generated a weighted sample $\mathcal{I}_i$ of \{$\theta,M_z\}$ using RIFT, where weights were proportional to the likelihood of the sample. For a given sample $j\in \mathcal{I}_i$ we generated the corresponding simulated waveform $h_{ij}$. This sample was compared to the observed gravitational wave. As the real incoming gravitational waveform is not known, we approximated it by its reconstructed waveform $h^{\rm rec}_{\rm gw}$. This reconstruction from the recorded data was carried out through identifying the coherent signal in the two LIGO detectors' data using the model-independent reconstruction algorithm cWB. In Fig. \ref{fig:time-domain} we show $\hat{h}^{\rm rec}_{\rm gw}$ for GW190521 in comparison with our highest-significance numerical relativity simulation, where $\hat{h}$ indicates spectral whitening. 

We quantified the similarity between $h_{ij}$ and $h^{\rm rec}_{\rm gw}$ with the normalized+whitened cross-correlation 
\begin{equation}
r_{\rm c}(h_{ij},h^{\rm rec}_{\rm gw}) = \frac{(\hat{h}_{ij} \star \hat{h}^{\rm rec}_{\rm gw})}{\sqrt{(\hat{h}_{ij} \star \hat{h}_{ij})(\hat{h}^{\rm rec}_{\rm gw} \star \hat{h}^{\rm rec}_{\rm gw})}}\,,
\label{eq:cc}
\end{equation}
For a waveform that perfectly matches the incoming gravitational wave and in the presence of no noise, we would get $r_{\rm c}=1$. 
%
The distribution of $r_{\rm c}(h_{ij},h^{\rm rec}_{\rm gw})$ over the weighted samples for our highest-likelihood numerical relativity simulation and, for comparison, highest-significance non-eccentric waveform, are shown in Fig.~\ref{fig:time-domain} (c,f--red histograms). 

To quantify how much of the measured similarity can be due to detector noise, 
we computed the distribution \addd{$r_{\rm c}(h_{ij}, h^{\rm rec}_{ij})$}. This compares each waveform to itself in the presence of noise, i.e. all deviation from $r_{\rm c}=1$ is due to the noise. We generated $h^{\rm rec}_{ij}$ by superimposing the waveform on real LIGO data. We show such control distributions in Fig.~\ref{fig:time-domain} (c,f--gray histograms). 

We quantified how well a simulated waveform (with a weighted sample of $M_z$ and extrinsic parameters $\theta$) matches $h^{\rm rec}_{\rm gw}$ by computing the probability $p_{\rm gw}$ that $r_{\rm c}(h_{\rm ij},h^{\rm rec}_{,gw})>r_{\rm c}(h_{ij},h^{\rm rec}_{,\it ij})$ for randomly and independently drawn $j,k\in \mathcal{I}_i$ (see Methods). \addd{We considered waveforms with $p_{\rm gw}>10^{-2}$ to pass the consistency test. Failing this criterion means that the waveform is ruled out with $99\%$ confidence; however, this test by itself does not establish any waveform as the correct explanation of the data. In Fig. \ref{fig:LnL} we indicated which waveform passed the consistency test.}

\section{Reconstructed binary parameters} 

We reconstructed the binary parameters over our numerical waveforms, weighing each waveform with their likelihood. We used a uniform prior on total mass, spin direction and spin magnitude, and eccentricity. The estimated parameters may depend on these choices\cite{2021ApJ...907L...9N} (see Supplementary Figs. 3, 4 \& \addd{6}). We list the reconstructed binary parameters found by our analysis in Table \ref{table:PE}, and for comparison the parameters found by LIGO-Virgo using a non-eccentric model\cite{2020arXiv201014527A}. While our results mainly constrain the masses, distance and eccentricity, we gain little information on the mass ratio, and the credible interval of our obtained precessing spin covers almost the full $[0,1]$ range, therefore our results effectively do not constrain precession.


We characterized the evidence supporting high eccentricity by comparing the likelihood values for our best-matching $e>0.5$ (${\cal L}_{\rm marg,high}$) and the best-matching $e<0.5$ (${\cal L}_{\rm marg,low}$) numerical relativity simulations.  This comparison supports $e>0.5$ with a Bayes factor of $\mathcal{B}={\cal L}_{\rm marg,high}/{\cal L}_{\rm marg,low}\approx76$. 
To bound the effect of limited sampling, we additionally computed ${\cal L}_{\rm marg}$ with the same technique as above, but using the full NRSur7dq4 family of non-eccentric, precessing waveforms that were used by LIGO-Virgo\cite{GW190521discovery,GW190521_properties} (see also Supplementary Fig. 1). With this comparison we obtained a Bayes factor of $\mathcal{B}\approx24$ in favor of the eccentric case over the NRSur7dq4 model. \addd{We additionally carried out waveform injection and reconstruction studies that confirmed that our method correctly identifies high- and zero-eccentricity waveforms (see Methods).}

\adddd{As an alternative to the likelihood comparison, we computed the maximum confidence interval over the eccentricity parameter space that excludes our $e=0$ results. We found that zero eccentricity can be excluded with $90$\% confidence.}

We found that the sensitive volume---the comoving volume within which a binary merger could have been detected---is a factor of two greater for our reconstructed model than in the non-eccentric case (see Methods). This means that, for identical rate densities, the eccentric case would be observed twice as often.

\section{Discussion} 

The 611 eccentric waveforms (effectively $\sim10^5$ templates) reported in this work enabled for the first time the probe of high eccentricity and precession in a black hole merger. \addd{While we evaluated GW190521, the presented method is readily applicable to other black hole mergers with total mass $\gtrsim50$\,M$_\odot$. The presented numerical relativity waveforms can also be the basis of probing low-mass binaries using hybrid waveforms\cite{2020PhRvD.102b4012S}, as highly accurate early inspiral waveforms can be obtained using semi-analytical approaches.}

High eccentricity at LIGO-Virgo's detectable frequencies can occur due to dynamical interactions in galactic nuclei\cite{2018ApJ...860....5G} or globular clusters\cite{2018PhRvD..97j3014S, 2018PhRvD..98l3005R}, due to Lidov-Kozai oscillations in isolated triple systems\cite{2017ApJ...836...39S,2019ApJ...881...41L} or near supermassive black holes\cite{2018ApJ...856..140H}, binary-single interactions in the disks of active galactic nuclei\cite{samsing2020active} (AGNs), and possibly during the merger of primordial black holes\cite{2016PhRvD..94h4013C}. For many of these scenarios\cite{2018PhRvD..97j3014S,2018PhRvD..98l3005R,2019ApJ...881...41L} a few percent of the black hole mergers could be highly eccentric with $e>0.5$, while for mergers in AGNs\cite{samsing2020active} and gravitational capture in galactic nuclei\cite{2018ApJ...860....5G} this fraction can be $\gtrsim 15\%$ and $\gtrsim 30\%$, respectively. Nonetheless, model uncertainties remain and even defining eccentricity at $e>0.5$ and high black hole masses is difficult (see Methods). 

The statistical significance of high eccentricity found here for GW190521 depends not just on our obtained Bayes factor, but also on the prior probability of a detected black hole merger being highly eccentric ($e>0.5$). For our fiducial analysis, an odds ratio of 10 in favor of an eccentric binary requires that at least 13\% of mergers are highly eccentric. The fraction of LIGO-Virgo's detections that are highly eccentric are likely below this fraction. For example, AGN-assisted mergers may contribute about 25\% of LIGO-Virgo's observations\cite{gayathri2021black}, while 30\% of these mergers could be highly eccentric\cite{samsing2020active}. For these fiducial values, $\sim8\%$ of LIGO-Virgo's detections could be highly eccentric. The fraction of highly eccentric mergers might nevertheless be higher among GW190521-like mergers. Let us consider the possibility that the high mass of GW190521 points to its dynamical origin. The fraction of highly eccentric binaries in such encounters ranges from $\sim5$\%\cite{2018PhRvD..97j3014S,2018PhRvD..98l3005R} to $\sim50$\%\cite{2018ApJ...860....5G}, i.e. could exceed $13\%$. While these estimates are illustrative, the currently available limited observational information on binary black hole systems and their formation mechanisms make our odds ratio estimates uncertain, precluding firm conclusions.

A possible astrophysical scenario that can lead to both high eccentricity and high mass is a hierarchical black hole merger that is the product of multiple previous chance encounters. Chance encounters can naturally lead to an eccentric merger, while multiple mergers can increase the black hole mass to beyond what is achievable through stellar evolution. Gravitational capture in a chance encounter will bring together two black holes with randomly oriented spins, with the black hole spins typically misaligned from the binary orbit, resulting in high $\chi_{\rm p}$. In the AGN scenario, misaligned spins are the norm due to the anisotropic binary-single interactions that also produce the high eccentricity\cite{samsing2020active}. However, the reconstructed mass and spin of the black holes alone are not sufficient to point to the origin of GW190521, and may also be consistent with an isolated binary origin. The observed high masses may be achievable directly through stellar core collapse, while high spin is possible in isolated binaries through tidal interactions\cite{2020A&A...636A.104B}. Precessing spin, which is uncommon in isolated binaries\cite{2015PhRvD..92f4016G}, is not well constrained. Eccentricity, therefore, could be a clearer signature of GW190521's possible origin.

\section{Methods}

\subsection{Numerical Relativity Simulations}

We carried out numerical relativity simulations of eccentric black hole mergers using the {\sc
LazEv} code~\cite{Zlochower:2005bj}, an implementation of the moving puncture approach~\cite{Campanelli:2005dd}. We used the BSSNOK (Baumgarte-Shapiro-Shibata-Nakamura-Oohara-Kojima) family
of evolution systems~\cite{Nakamura87, Shibata95, Baumgarte99}.
All simulations used 6$^{th}$ order spatial finite-differencing, 
5$^{th}$ order Kreiss-Oliger dissipation, and Courant factors of 1/3~\cite{Lousto:2007rj,Zlochower:2012fk,Healy:2016lce}.
These techniques are the same used to generate quasicircular simulations~\cite{Healy:2017psd,Healy:2019jyf,Healy:2020vre}, as our formalism is robust to deal with generic orbits.

The {\sc LazEv} code uses the {\sc EinsteinToolkit}~\cite{Loffler:2011ay, einsteintoolkit} / {\sc Cactus}~\cite{cactus_web} / {\sc Carpet}~\cite{Schnetter-etal-03b} infrastructure.  The {\sc Carpet} mesh refinement driver provides a ``moving boxes'' style of mesh refinement. In this approach, refined grids of fixed size are arranged about the coordinate centers of both holes.  The code then moves these fine grids about the computational domain by following the trajectories of the two black holes.

To compute the numerical initial data, we used the puncture
approach~\cite{Brandt97b} along with the {\sc  TwoPunctures}
~\cite{Ansorg:2004ds} code.  For each eccentric family, we 
first determined the initial separation and tangential quasicircular momentum, $p_{t,qc} $, that leads to a quasicircular frequency of 10\,Hz for a $50\,$M$_\odot$ system, using the
post-Newtonian techniques described in~\cite{Healy:2017zqj}.  
To increase the eccentricity of the system while keeping the initial data 
at an apocenter, the initial tangential momentum was modified by parameter, $0 < \epsilon < 1$,
such that $p_t = p_{t,qc} ( 1 - \epsilon )$.  The eccentricity was then approximately
$e = 2\epsilon-\epsilon^2$ and the initial frequency $10$\,Hz$( 1 - \epsilon )$. With this procedure the initial mean anomaly of the system is 180$^{\circ}$.

Gravitational waveforms were calculated via the radiative Weyl
Scalar $\psi_4$, which was decomposed into $\ell$ and $m$ modes.  
We extracted the gravitational radiation at
finite radius and extrapolated to $r=\infty$ using the
perturbative extrapolation described in Ref.~\cite{Nakano:2015pta}.
The gravitational strain waveform was then calculated from the
extrapolated $\psi_4$ in Fourier space\cite{Campanelli:2008nk,Reisswig:2010di}.

We carried out $611$ eccentric binary black hole simulations in this study, with eccentricities in the full $e \in (0,1]$ range. These simulations included non-spinning ($313$), aligned-spin ($37$), antialigned-spin ($123$), head-on ($35$), and spin-precessing (111) waveforms, and mass ratios $1/7 \leq q=m_2/m_1 \leq 1$. We first carried out the most thorough survey of the eccentricity-mass ratio parameter space with non-spinning simulations. Then we carried out aligned/anti-aligned spin and precessing simulations for a broad range of eccentricity values, most densely covering the parts of the parameter space where the non-spinning simulations produced the highest $\log{\cal L}_{\rm marg}$ in comparison to GW190521. For precessing waveforms we most densely targeted the $\chi_{\rm p}\sim 0.7$ case expected from black hole merger remnants.

In addition to these eccentric simulations in the presented analysis we used an additional $920$ numerical relativity simulations with $e=0$. These waveforms cover essentially the entire plausible parameter space with a mass ratio range of $q\in[0.015,1]$ effective spin range of $\chi_{\rm eff}\in[-0.98,0.98]$, and precessing spin range of $\chi_{\rm p}\in[0,0.85]$ (see also Supplementary Table 2).

Our simulated gravitational waveforms were then scaled to correspond to different black hole masses, providing about $100$ scaled mass values for each simulation. 
In our Supplementary Table 1 we present a detailed list of the parameters selected for each of our simulations, along with the computed $\log{\cal L}_{\rm marg}$, the result of the cWB consistency check and the reconstructed distance of the source given the simulated parameters. For black hole spins, Supplementary Table 1 lists the binary's effective spin $\chi_{\rm eff}= c G^{-1} (m_1+m_2)^{-1}(\vec{S}_1/m_1+\vec{S}_2/m_2)\vec{L}/|\vec{L}|$, which describes the objects' spin component parallel to the binary's orbital axis, and precessing spin $\chi_{\rm p}$=$\max \left(\vec{S}_1/m_1,\frac{2+3q/2}{q^2} \vec{S}_2/m_2\right)\times \vec{L}/|\vec{L}|$ that describes the projection of component
spin vectors perpendicular to the orbital axis\cite{PhysRevLett.113.151101}. For eccentricity, Supplementary Table 1 lists the binary's eccentricity at 10\,Hz assuming a binary mass of $50$\,M$_{\odot}$. For our simulations starting at separations of $\sim 25M$ (with $c=G=1$ units), the small eccentricity waveforms take many orbits to merge, but those for high eccentricity ($e\gtrsim0.5$) are an almost direct plunge (see Supplementary Fig. 2), hence we provide the initial eccentricity value.

The evolution techniques used in our simulations have been validated by convergence studies, comparisons with Post-Newtonian waveforms and perturbations in their appropriate regimes\cite{2009PhRvD..79h4010C,2010PhRvL.104u1101L}. We carried out cross checks against independent numerical relativity simulations by other groups, for example in the case of gravitational wave events GW150914\cite{2016CQGra..33x4002L}, GW170104\cite{2018PhRvD..97f4027H} and other events\cite{2020PhRvD.102l4053H} in mode per mode convergence studies. While these validations were carried out with $e\ll1$, there is no particular loss of accuracy in the plunging regime in our full numerical approach. In addition, we carried out a study that generated an eccentric waveform with $e=0.7$ at 3 different spatio-temporal resolution levels, with the coarsest resolution being our baseline run, while the other two having a factor of 1.2 and 1.44 times higher resolution. We confirmed that both higher spatio-temporal resolutions yield consistent $\log({\cal L}_{\rm marg})$ and reconstructed binary parameter values with our baseline run.

\subsection{Waveform consistency computation}


We estimated the probability density of the binary parameters corresponding to the \addd{observed gravitational wave based on numerical relativity simulation $i$}, using the RIFT package. The simulated waveform $i$ has intrinsic parameters including the black holes' mass ratio $q$, spins $\vec{S}_1$ and $\vec{S}_2$ and binary eccentricity $e$. These are fixed in this parameter estimation. Instead, estimation is over the extrinsic parameters $\theta$ (luminosity distance, time, two sky position angles, and three Euler angles describing the source orientation) and the redshifted total mass $M_{\rm z}$. The RIFT estimation returns a Monte Carlo-sampled ensemble of sets of binary parameters, denoted with $\mathcal{I}_i$, where the density of the selected parameters corresponds to the reconstructed probability density of the binary parameters. In the present study we generated $N_{\mathcal{I}_i}=5000$ such samples for each estimation. For a set of parameters $\{\theta,M_{\rm z}\}_j\in \mathcal{I}_i$ \addd{corresponding to one sample,} we superimposed the corresponding gravitational waveform, denoted with $h_{ij}$, with \addd{real} gravitational-wave detector noise, and reconstructed its waveform using cWB, obtaining reconstructed waveform $h^{\rm rec}_{ij}$. \addd{cWB coherently combines information from all detectors while also removing some characteristic non-Gaussian detector noise features.} 
The same procedure was repeated for each waveform within $\mathcal{I}_i$. We then computed the probability
\addd{
\begin{equation}
p_{\rm gw,i} = \frac{2}{N}\sum \underset{j\in \mathcal{I}_i}{\mbox{random}}\left(H\left[ r_{\rm c}(h_{ij},h^{\rm rec}_{\rm gw}) - r_{\rm c}(h_{ij},h^{\rm rec}_{ij}) \right]\right)
\end{equation}}
where \addd{$H$ is the Heaviside step function, "random" denotes the random selection of two sets of parameters from $\mathcal{I}_i$, and the sum runs over $N=10^4$ random selections. This is the probability that a randomly selected $h_{ij}$ waveform will be more similar to $h^{\rm rec}_{\rm gw}$ than to its own reconstructed waveform $h^{\rm rec}_{ij}$. If a given $h_{ij}$ is the "correct" waveform describing the incoming gravitational wave then the probability density of $r_{\rm c}(h_{ij},h^{\rm rec}_{\rm gw})$ should follow that of $r_{\rm c}(h_{ij},h^{\rm rec}_{ij})$, i.e. such the presence of such a waveform within $\mathcal{I}_i$ should "increase" $p_{\rm gw}$. Therefore, $p_{\rm gw}$ may be thought of as the fraction of waveforms within $\mathcal{I}_i$ for which $r_{\rm c}(h_{ij},h^{\rm rec}_{\rm gw})<1$ effectively only due to the presence of detector noise, and not because of intrinsic mismatch between $h_{ij}$ and the incoming gravitational wave.}

While low $p_{\rm gw}$ indicates that the waveform is unlikely to be a good explanation of the observed data, a high $p_{\rm gw}$ can occur not just for accurate waveforms, but also in the presence of significant noise as a low signal-to-noise ratio signal can broaden the $r_{\rm c}(h_{i},h^{\rm rec}_{i})$ distribution. For example in Fig. \ref{fig:LnL} we see that a few of the low-likelihood events pass the consistency test. These in general are waveforms reconstructed with relatively low signal-to-noise ratio.

\subsection{Sensitive volume computation}

For the sensitive volume computation, we considered our maximum likelihood numerical relativity simulation. We carried out a Monte Carlo simulation in which each each realization was a randomly generated gravitational wave signal. The signal was selected to have a random sky location drawn isotropically, and random distance assuming uniform distribution in comoving-volume. The maximum distance for the injections was chosen to be greater than where the signal can be realistically detected. The binary's total mass (in the source frame) was randomly selected from the reconstructed mass estimates generated by RIFT for the maximum-likelihood waveform, which reflect the estimated mass probability density. We superimposed each generated waveform with gravitational-wave data from the two LIGO observatories, and executed the cWB algorithm to detect them (with false-alarm-rate threshold of $0.33$\,yr$^{-1}$). We computed the sensitive volume by taking the ratio of the detected and generated waveforms, multiplied by the volume in which the signals were generated.

Other than eccentricity, the obtained sensitive volume mainly depends on the total binary mass. We used a standard uniform prior mass distribution for this analysis, however the underlying black hole mass distribution is likely different from this that could affect our mass estimate\cite{2020ApJ...891L..31F}.

\section{Robustness of eccentricity reconstruction} 

LIGO-Virgo carried out a binary parameter estimation for GW190521 and found mild evidence for spin-induced orbital precession\cite{GW190521discovery}. The waveform family used for that analysis, however, did not include the possibility of eccentricity, and in general it is difficult to identify orbital eccentricity through observations\cite{Abbott_eBBH,Ramos-Buades:2020eju,2013PhRvD..87d3004E}, e.g. since highly-eccentric, high-mass black hole mergers may be confused with precession if one uses a non-eccentric, precessing waveform model\cite{bustillo2020confusing}. For GW190521, low-eccentricity, non-precessing waveforms match observations similarly to precessing but non-eccentric waveforms\cite{IsobelGW190521,Lower_2018}. 

The NRSur7dq4 model densely covers the non-eccentric parameter space, while our numerical relativity simulations have a sparser coverage. The Bayes factor comparing our high-eccentricity result with the best-matching NRSur7dq4 case can be considered as a lower limit on the Bayes factor for a complete eccentric waveform family. A caveat with this comparison is that the NRSur7dq4 model only includes mass ratios of $q>0.17$. Several recent analyses raised the possibility of GW190521 having a smaller mass ratio of $q\sim0.1$.\cite{2021arXiv210506360E,2021arXiv210509151N,2020ApJ...904L..26F}, which is not accounted for in our Bayes factor. Our non-eccentric simulations extend down to $q\approx0.015$, although they do not densely cover this space.

As a further sanity check of our analysis technique and results, we carried out three injection studies to evaluate the outcome of the analysis on known gravitational waveforms. In these we superimposed real LIGO-Virgo recorded data with simulated gravitational waveforms, and carried out the full reconstruction process identically to how it was done for GW190521. The first study used a numerical relativity simulation with $e=0.75$, assuming binary mass and spin comparable to our best-matching waveform. The simulated waveform we adopted was not included in the waveform family used for the reconstruction. Our analysis recovered a binary eccentricity of $e=0.71\pm0.1$ ($1\sigma$ error), and found a statistical support for $e>0.5$ over $e<0.5$ of $\mathcal{B}\approx73$. The second study used a numerical relativity simulation with $e=0$, assuming binary properties similar to those reconstructed by LIGO-Virgo. Our analysis recovered $e=0.04\pm0.1$ ($1\sigma$ error), and $\mathcal{B}\approx58$ supporting $e<0.5$ over $e>0.5$. In a third study we injected a waveform from the NRSur7dq4 family ($e=0$) to probe the role of a potential difference between these waveforms and our numerical relativity simulations. For this injection our analysis recovered $e=0.04\pm0.15$ ($1\sigma$ error), and $\mathcal{B}\approx10$ supporting $e<0.5$ over $e>0.5$. For all three cases, we see that the eccentricity identification is reasonably accurate, and the distinction between low and high-eccentricity cases is statistically significant.

Our results on significance are dependent on the assumed prior distributions of the binary parameters. In particular, highly eccentric mergers are expected to be less common in the universe than non-eccentric events. Considering our numerical-relativity comparison that obtains a Bayes factor of $\mathcal{B}\approx76$ and a threshold of $\mathcal{B}_{\rm th}=10$ for indication of eccentricity, one would need the expected fraction of $e>0.5$ binaries to be at least $13\%$.


\subsection{Data Availability}

Numerical relativity waveforms generated for and used in this study will be accessible at \url{http://ccrg.rit.edu/~RITCatalog}. Data generated by our calculations are available in the Supplementary material or are available from the corresponding authors upon reasonable request.

\noindent{\bf Code availability} The computer code that was used for the calculations is available from the corresponding authors upon reasonable request.

\noindent{\bf Acknowledgements}
\noindent The authors would like to thank Chris Belczynski, Erik Katsavounidis, Avi Loeb and Salvatore Vitale for useful suggestions, and Gabriele Vedovato for his valuable help with the cWB search algorithm and with the setup of the cWB waveform consistency study. The authors gratefully acknowledge the National Science Foundation (NSF)
for financial support from Grants
No.\ PHY-1912632 (C.L., M.C., R.O.), No.\ PHY-1911796 (I.B.), No.\ PHY-1806165 (S.K.), No.\ PHY-1707946 (M.C.), No.\ ACI-1550436 (M.C.), No.\ AST-1516150 (M.C.),
No.\ ACI-1516125 (M.C.), No.\ PHY-1726215 (M.C., C.L.), NASA TCAN grant No. 80NSSC18K1488 (M.C.,R.O.), and the support of the Alfred P. Sloan Foundation (I.B.).
This work used the Extreme Science and Engineering Discovery Environment (XSEDE) [allocation TG-PHY060027N], which is supported by NSF grant No. ACI-1548562 and Frontera projects PHY-20010 and PHY-20007.
Computational resources were also provided by the NewHorizons, BlueSky Clusters, and Green Prairies at the Rochester Institute of Technology, which were supported by NSF grants No.\ PHY-0722703 (M.C., C.L.), No.\ DMS-0820923, No.\ AST-1028087 (M.C.), No.\ PHY-1229173 (M.C., C.L.), and No.\ PHY-1726215 (M.C., C.L.). We are grateful for computational resources provided by the Leonard E Parker Center for Gravitation, Cosmology and Astrophysics at the University of Wisconsin-Milwaukee supported by NSF Grant No. PHY-1626190. We also acknowledge the use of IUCAA LDG cluster Sarathi for the 
computational/numerical work. We also acknowledge the use of LDG clusters at CIT, LHO, and LLO for the computational/numerical work.
This research has made
use of data, software and/or web tools obtained from
the Gravitational Wave Open Science Center (https:
//www.gw-openscience.org), a service of LIGO Laboratory, the LIGO Scientific Collaboration and the Virgo
Collaboration.
LIGO is funded by the U.S. National Science Foundation. Virgo is funded by the French Centre National de Recherche Scientifique (CNRS), the Italian
Istituto Nazionale della Fisica Nucleare (INFN) and the
Dutch Nikhef, with contributions by Polish and Hungarian institutes.

\noindent{\bf Author contributions}
\noindent I.B and R.O. contributed to the origination of the idea. J.H., J.L., C.L., M.C. and R.O. carried out the numerical relativity simulations and the RIT sensitivity studies. V.G., B.O., M.S., S.K. and I.B. carried out the cWB consistency check. All authors worked out collaboratively the general details of the project. V.G. created the figures with input from all other authors. All
authors helped edit the manuscript.

\noindent{\bf Competing interests}
The authors declare no competing interests.

\noindent{\bf Additional information}
Correspondence and requests for materials should be addressed to I.B.~(email: imrebartos@ufl.edu) and R.O~(email: rossma@rit.edu).

\begin{table}
\renewcommand{\arraystretch}{0.8}
\centering
\begin{tabular}{lrr}
\hline
\hline
Parameters & \,\,\,\,\,\,\,\,\,\,This work & \,\,\,\,\,\,\,\,\,\,\,LIGO-Virgo \\
\hline
Primary mass $[$M$_\odot]$       & $78^{+ 9 }_{ -5 }$ & $95^{+29}_{-19}$\\
Secondary mass $[$M$_\odot]$      & $78^{+ 9 }_{ -5 }$ & $69^{+22}_{-24}$ \\
Total mass $[$M$_\odot]$          & $155 ^{+ 17 }_{ -11 }$ & $164^{+39}_{-23}$ \\
Mass ratio$^\star$                        & $1$ & $0.75^{+0.22}_{-0.35}$ \\
Luminosity distance $[$Gpc$]$     & $7.7 ^{+ 1.27 }_{ -1.65 }$ & $3.9^{+2.2}_{-2.0}$ \\
Redshift                          & $1.13 ^{+ 0.15 }_{ -0.2 }$ & $0.68^{+0.28}_{-0.28}$  \\
Eccentricity                      & $0.69^{+0.17}_{-0.22}$ & $0$ \\
Effective spin $(\chi_{\rm eff})$ & $0.27^{+0.56}_{-0.51}$ & $0.03^{+0.31}_{-0.40}$ \\
Precession spin $(\chi_{\rm p})$  & $0.66^{+0.28}_{-0.60}$ & $0.67^{+0.26}_{-0.44}$ \\
Sensitive volume$^\dagger$ $[$Gpc$^{3}]$      & $46.5$ & $25.3$  \\
\hline
\hline
\end{tabular}
\caption{{\bf Reconstructed properties of GW190521.} For comparison we also show the properties obtained by LIGO-Virgo\cite{2020arXiv201014527A} using the NRSur7dq4 non-eccentric, precessing waveform model\cite{PhysRevResearch.1.033015}. Error bars show 90\% credible intervals. $^\star$Reported here for maximum-likelihood waveform (see \cite{2021arXiv210506360E,2020ApJ...904L..26F,2021ApJ...907L...9N} for the possibility of GW190521 having a low mass ratio). $^\dagger$Computed for the maximum-likelihood waveform, marginalizing over total mass (see Methods).}
\label{table:PE}
\end{table}

\newpage
\bibliography{Refs}

\end{document}